%% file: iclr2025_conference.tex
\documentclass{article} 
\usepackage{iclr2025_conference,times}

\input{math_commands.tex}

\usepackage{hyperref}
\usepackage{url}
\usepackage{longtable} 
\usepackage{graphicx} 
\usepackage{array} 
\usepackage{booktabs}
\usepackage{pdflscape}
\usepackage{lscape}
\usepackage{makecell}

\title{Compressed code: the hidden effects of quantization and distillation on programming tokens
}

\author{Viacheslav Siniaev \\
National Sun Yat-Sen University \\
Kaohsiung, Taiwan \\
\texttt{viacheslavsiniaev@gmail.com} \\
\And
Iaroslav Chelombitko \\
DataSpike; \\
aglabx; \\ 
Neapolis University Pafos \\
Pafos, Cyprus \\
\texttt{i.chelombitko@nup.ac.cy} \\
\And
Aleksey Komissarov \\
aglabx; \\
Neapolis University Pafos \\
Pafos, Cyprus \\
\texttt{ad3002@gmail.com}
}

\iclrfinalcopy 
\begin{document}

\maketitle

\begin{abstract}

Large Language Models (LLMs) have demonstrated exceptional code generation capabilities, yet their token-level mechanisms remain underexplored, particularly in compressed models. Through systematic analysis of programming language token representations, we characterize how programming languages are encoded in LLM tokenizers by analyzing their vocabulary distribution and keyword coverage patterns. We introduce a novel cold-start probability analysis method that provides insights into model behavior without requiring explicit prompts. Additionally, we present a comprehensive evaluation of how different model optimization techniques - including quantization, distillation, model scaling, and task-specific fine-tuning - affect token-level representations and code generation quality. Our experiments, supported by comprehensive probability distribution analysis and evaluation metrics, reveal critical insights into token-level behavior and provide empirically-validated guidelines for maintaining code generation quality under various optimization constraints. These findings advance both theoretical understanding of LLM code generation and practical implementation of optimized models in production environments.

The reproducible code for our work is available on GitHub: \url{https://github.com/Egelvein/compressed_code}.

\end{abstract}

\section{Introduction}
\subsection{Motivation}
Programming language understanding and code generation are critical applications of LLMs. Tokenization plays a crucial role in shaping model efficiency and performance. Byte Pair Encoding \citep{bostrom2020byte} is widely used but often introduce inefficiencies, particularly in programming languages where syntax-specific tokens must be preserved. Poor tokenization can lead to increased sequence length, higher computational cost, and reduced model interpretability.


While LLMs excel at natural language tasks, their ability to handle programming languages remains dependent on how well their tokenizers preserve syntax and semantics. Many tokenizers are optimized for general text but fail to efficiently capture code structure, leading to suboptimal code completion and generation results (see \citealt{dagan2024tokenizer}). Recent work has shown that tokenizer quality significantly impacts model performance across different languages and domains (\citealt{Chelombitko2024}), with specialized tokenizers demonstrating substantial improvements in compression ratios and representation efficiency (see \citealt{chelombitko-komissarov-2024-specialized}). An emerging alternative 
direction explores tokenizer-free architectures: the Meta-AI Byte Latent Transformer 
(see \citealt{pagnoni2024bytelatenttransformerpatches}) demonstrates that byte-level models can match 
tokenization-based LLM performance at scale, suggesting that the relationship 
between tokenization and model capabilities may be more nuanced than previously assumed.

Industry adoption of LLM compression techniques such as quantization and distillation is rapidly increasing, with over 73\% of enterprise deployments relying on compressed models. However, little research has been conducted on how compression impacts programming language tokenization. Our study aims to fill this gap by systematically analyzing the effects of compression on token representation and model performance.

Efficient model deployment requires balancing compression techniques with maintaining generation quality. Understanding how token probability distributions shift due to compression will provide insights into ensuring reliable code completion in real-world applications.

\subsection{Background and Related Work}

Recent research has focused on improving LLM performance for natural language tasks, yet few studies have delved into token-level behavior in code-specific contexts. Prior work on model compression has largely targeted natural language tasks, leaving a gap in our understanding of its effects on programming language tokens. Our work builds on and extends these findings by examining token vocabulary composition, generation probability distributions, and model confidence metrics specifically for programming languages.

Despite these advances, few studies delve deeply into the granularity of token representations for programming languages. Existing research often assesses model performance via downstream tasks using traditional benchmarks like HumanEval \citep{chen2021evaluating} or MBPP \citep{austin2021program}. However, the shift towards evaluating complex instruction following and repository-level reasoning exemplified by BigCodeBench \citep{zhuo2025bigcodebenchbenchmarkingcodegeneration} and SWE-bench \citep{jimenez2024swebenchlanguagemodelsresolve} - highlights the need for deeper interpretability. These modern benchmarks reveal that high pass rates do not always correlate with structural understanding, yet few studies link these performance metrics to the granularity of token representations or probability evolution.


Common model compression strategies include quantization (reducing numerical precision), distillation (teacher-student paradigms), and pruning (removing weights or neurons). While compression methods are widely studied for natural language tasks, code-focused compression remains underexplored. Preliminary evidence suggests that quantization can degrade the ability to handle rare tokens \citep{ouyang2024lowbit}, and distillation can lead to biased distributions if specialized tokens are not preserved.

By examining token probabilities pre- and post-compression, our work provides a clearer picture of why and where code completion might fail in constrained environments.

\subsection{Our Contribution}

Our primary contributions are threefold. First, we introduce a novel methodology for evaluating code-specific model capabilities through token probability analysis, including our "cold start" metric that provides insights into model behavior without requiring extensive testing or benchmarking. This approach offers a new perspective on how models represent and process programming constructs, moving beyond traditional accuracy-based evaluations.

Second, we present surprising findings about the nature of code-specialized models, demonstrating that they do not develop fundamentally different token distributions compared to general language models. This insight challenges common assumptions about model specialization and suggests that the relationship between tokenization and code generation capabilities is more nuanced than previously understood.

Third, we provide the first systematic analysis of how compression techniques affect code generation capabilities at the token level. Our results reveal unexpected non-linear effects, where moderate quantization can actually improve the balance between programming keywords and special tokens, while distillation can significantly alter token probability distributions. These findings have immediate practical implications for deploying compressed models in production environments while maintaining code generation quality.

This research bridges the gap between theoretical understanding of language models and practical deployment considerations, offering both novel analytical tools and actionable insights for maintaining reliable code generation capabilities under resource constraints. 

\section{Methodology and Results}

\subsection{Choosing coding models and tokenizers}

As a starting point in our study, we selected a set of top-performing open-source language models that demonstrated the best performance in coding tasks while also being publicly available in the Chatbot Arena benchmark \citep{chiang2024chatbot}. The models included in our analysis are Qwen2.5-Coder, Qwen2.5 \citep{hui2024qwen25}, Athene-V2-Chat \citep{nexusflow2024athenev2}, DeepSeek-V2.5 \citep{deepseek2024deepseekv2}, DeepSeek-V3 \citep{deepseek2024deepseekv3}, DeepSeek-R1, and additionally DeepSeek-R1 distillations: DeepSeek-R1-Llama-70B, DeepSeek-R1-Llama-8B, DeepSeek-R1-Qwen-32B, DeepSeek-R1-Qwen-14B, DeepSeek-R1-Qwen-7B, DeepSeek-R1-Qwen-1.5B \citep{deepseek2025deepseekr1}, and Llama 3.1 \citep{grattafiori2024llama3}.

Upon analyzing the tokenizer vocabularies of the selected models, we identified several instances where different models share identical tokenization schemes. Specifically, we found that Qwen2.5-Coder, Qwen2.5, Athene-V2-Chat, DeepSeek-R1-Qwen-32B, DeepSeek-R1-Qwen-14B, DeepSeek-R1-Qwen-7B, and DeepSeek-R1-Qwen-1.5B use the same vocabulary. Similarly, DeepSeek-R1 and DeepSeek-V3 share an identical vocabulary, indicating continuity in tokenization strategies across different versions of the DeepSeek series. Finally, we observed that Llama 3.1, DeepSeek-R1-Llama-70B, and DeepSeek-R1-Llama-8B use the same tokenizer vocabulary, which aligns with expectations given their architectural similarities. Thus, only three unique tokenizers remain for further analysis, which we will focus on in our study. From now on, we will refer to the tokenizers by the names of their respective models.

\subsection{Keywords as language markers in tokenizers}

Next, we considered the most popular programming languages on GitHub \citep{githuttwo}, as this metric better reflects real-world usage and developer activity. Based on recent rankings, we selected the top ten programming languages: Python, Java, Go, JavaScript, C++, TypeScript, PHP, Ruby, C, and C\#. Additionally, we included Rust and React, the last is a widely used JavaScript library for building user interfaces, as it has gained significant traction in code generation tasks using large language models. To assess how well these languages and frameworks are represented in the tokenizers, we adopted the following approach. For each programming language, we compiled a set of its reserved keywords, while for React, we collected key framework-specific keywords. These sets were then used as a basis for evaluating their presence in the tokenizers. 

To further analyze how programming-related tokens are represented in three tokenizers, we compiled a dataset of reserved keywords from multiple programming languages (276 keywords) and examined their rankings. 

\begin{figure}[h]
    \centering
    \includegraphics[width=1.0\linewidth]{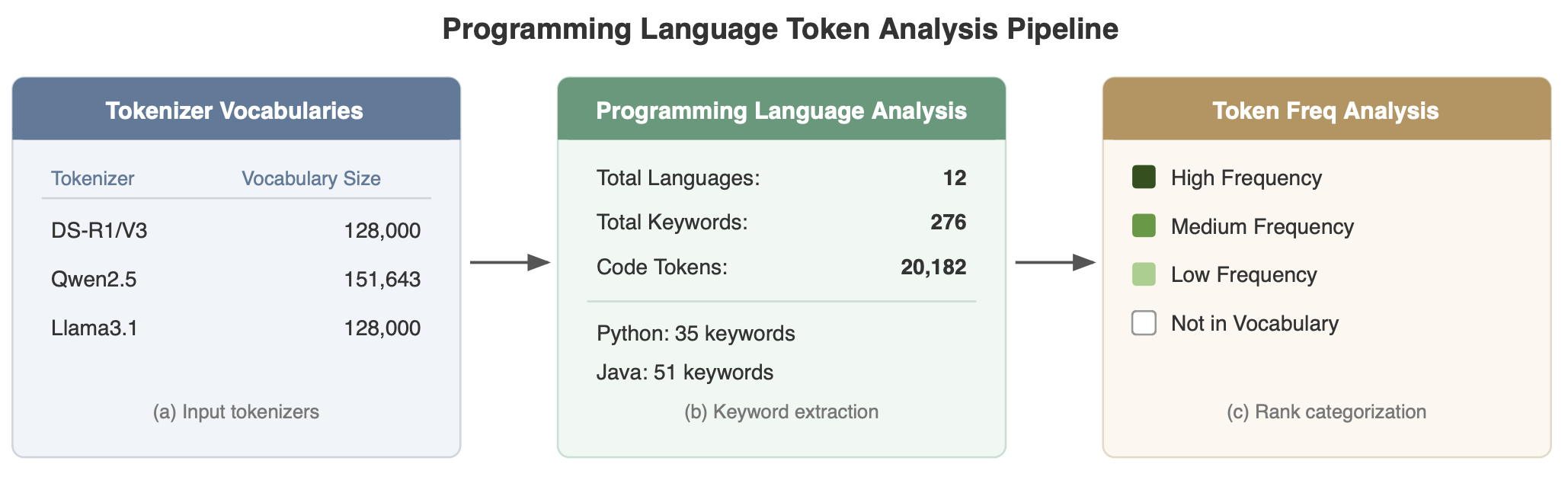}
    \caption{Pipeline for analyzing programming language tokens in LLM tokenizers across three major models (DeepSeek-R1/V3, Qwen2.5, and Llama 3.1). The process analyzes vocabulary composition (128K-151K tokens), examines keyword representation across 12 programming languages (276 unique keywords), and categorizes tokens by frequency to assess language coverage.}
    \label{fig:pipeline}
\end{figure}

\subsection{Vocabulary Analysis}

We aimed to determine how well language-specific tokens are represented in the tokenizers, following the process depicted in Figure~\ref{fig:pipeline}. This task can be divided into two parts. The first step was to identify which tokens should be considered language-specific. To address this, we selected tokens that correspond to reserved keywords in each programming language. The second step involved evaluating the rank of each token within the tokenizer’s vocabulary. For every token associated with a programming language, we examined its ranking to assess its prominence in the tokenizer. Appendix Tables~\ref{keyword-token-rank-analysis} and~\ref{keyword-coverage-comparison} provides an overview of the presence and relative ranking of these keywords across tokenizers. The ranking reflects how frequently a token appears within a tokenizer’s vocabulary, with lower ranks indicating higher frequency.

DeepSeek-R1 tokenizer consistently lags behind Qwen2.5-Coder and Llama 3.1 across all languages, likely because new tokens were added at the beginning of its vocabulary, shifting coding tokens to the right, or because its training dataset contained significantly less code than those used for Qwen and Llama 3.1. We also observe that modern, widely-used languages such as Python, TypeScript, Rust, Go, and C\# are well-represented, with keyword coverage exceeding 90\% in most cases. This aligns with their prevalence in contemporary software development and AI-assisted coding environments. In contrast, languages like C, C++, and React exhibit notably lower coverage, with React being the least represented. This suggests that these tokenizers may not be optimized for frameworks and lower-level system programming languages, potentially leading to inefficiencies in language model usage for these domains.

A key factor influencing token representation is the overlap between programming language keywords and natural language vocabulary. Languages like Python have many keywords that also exist in general vocabulary (e.g., return, in, with), making them more likely to be represented as single tokens due to their high frequency in pretraining corpora. In contrast, more domain-specific keywords, such as those in C, C++, and Rust, are rarely found in general text, leading to fragmentation in tokenization or complete absence from token vocabularies.

This observation suggests that languages with higher lexical overlap with natural language may be better understood by models, as their keywords are directly mapped to existing tokens. However, this can also introduce ambiguity, as these tokens were seen in different context during the model training. On the other hand, specialized programming keywords that do not intersect with natural language have the advantage of being unambiguous but may suffer from suboptimal tokenization, impacting model performance in tasks requiring precise syntactic understanding.

However, the mere presence or absence of a token in the tokenizer is not a sufficiently illustrative metric. A more informative measure is the ranking of the token within the tokenizer’s vocabulary. This ranking provides deeper insight into how different tokenizers prioritize language-specific tokens and how they distribute them within their vocabularies.

\subsection{Token ranking metric}

In addition to measuring keyword coverage, we analyzed the ranks assigned to programming language keywords in different tokenizers. Our findings reveal a systematic trend where DeepSeek-R1 assigns significantly higher ranks to keywords compared to Qwen2.5-Coder and Llama 3.1. Since rank in BPE-based tokenization is largely influenced by token frequency in the training corpus, this suggests that DeepSeek-R1 either has a different tokenization strategy or its training data contains fewer instances of these programming language keywords.

This discrepancy reinforces our earlier observation that DeepSeek-R1 is structurally different from the other tokenizers. A higher rank indicates that a keyword is tokenized less efficiently, meaning it may be split into multiple subword tokens rather than appearing as a single unit. This could impact model performance, particularly for programming tasks where tokenization granularity plays a crucial role in syntax comprehension. The results suggest that DeepSeek-R1 may not be as optimized for code-related tasks as Qwen2.5-Coder and Llama 3.1, which assign lower ranks to programming keywords and thus likely tokenize them more efficiently.

Additionally we checked the presence specific not word-releated tokens (Appendix Table~\ref{special-chars-coverage}). Llama 3.1 and Qwen2.5 have a significantly higher proportion of tokens containing programming-related symbols compared to DeepSeek-V3. This suggests that DeepSeek-V3 may have a more compressed or selective representation of syntax-related tokens, potentially impacting its efficiency in parsing and generating code.

One unexpected result was that we initially assumed coding-specific models would have distinct token distributions compared to general language models, with additional tokens specifically introduced during training to better represent programming languages. However, this turned out not to be the case. General language models and those designed for coding share the same set of tokens. This finding was surprising, as we expected programming-focused models to exhibit a tokenizer vocabulary more specialized for coding-related syntax and keywords.

Another unexpected finding is that a large portion of tokens in these tokenizers consists of special characters commonly used in programming and formatting rather than carrying direct semantic meaning. In Llama 3.1 - 14.6\% of the vocabulary is composed of such tokens, in Qwen2.5 it is 12.1\%, and in DeepSeek-V3 it is 5.1\%. This suggests that a considerable fraction of the tokenizer is allocated to structural elements—such as brackets, indentation, and operators—rather than words or identifiers. This distribution could impact how efficiently models process programming languages and structured data, as these non-semantic tokens play a crucial role in syntax representation but do not directly contribute to meaning in the traditional linguistic sense.

\subsection{A new metric to estimate coding abilities}
The next step in our analysis stemmed from the observation that the difference in tokens between models was not as significant as we initially expected. This can be explained by the fact that, despite analyzing multiple models, they all rely on only three distinct tokenizers. To gain further insight, we introduced an additional metric to evaluate how programming-related tokens are represented in pre-trained models. Specifically, we examined the probability of generating these tokens when no prompt is provided—a scenario we refer to as the cold start.
Under the cold start condition, we analyzed the individual probabilities of programming-related tokens appearing spontaneously, as well as their cumulative probability across different models. However, we were only able to compute this metric for the following models: DeepSeek-R1-Distill-Qwen-1.5B, DeepSeek-R1-Distill-Qwen-7B, DeepSeek-R1-Distill-Qwen-14B, DeepSeek-R1-Distill-Qwen-32B, DeepSeek-R1-Distill-Llama-8B, DeepSeek-R1-Distill-Llama-70B, Qwen2.5-Coder-7B-Instruct from DeepSeek R1 repository.
\subsubsection{Metrics Definition}
To comprehensively evaluate the models' coding capabilities, we introduced several key metrics:

\begin{enumerate}
    \item Programming Keywords Probability (PKP): This metric represents the cumulative probability of generating programming language keywords (e.g., 'def', 'import', 'class') in a cold start scenario. The PKP provides insight into the model's baseline tendency to generate actual programming constructs.
    \item Special Tokens Probability (STP): This metric measures the cumulative probability of generating special tokens commonly used in programming (e.g., parentheses, brackets, indentation markers). The STP helps understand the model's structural coding awareness.
    \item Keyword Average Probability (KAP): Calculated as the mean probability across all programming keywords, this metric allows us to assess the uniform distribution of keyword probabilities and identify potential biases toward specific programming constructs.
    \item Natural Language Probability (NLP): This serves as a control metric, measuring the probability of generating common natural language tokens. It helps establish a baseline for comparison with programming-specific probabilities.
\end{enumerate}

\subsubsection{Results Analysis}
Our analysis revealed several interesting patterns across the evaluated models (Appendix Table~\ref{tokenizer-comparison-1}). First, we observed that the Special Tokens Probability consistently dominated the probability distribution, with values ranging from 0.1006 to 0.7246 across different models. This suggests a strong structural bias in the models' architecture, favoring syntactic elements over semantic programming constructs.
The Programming Keywords Probability showed more variation between models, with Qwen2.5-Coder-7B-Instruct demonstrating notably higher values (0.0902) compared to other models in the DeepSeek family (0.0011-0.0062). This significant difference suggests that architectural choices and training objectives may have a stronger influence on coding capabilities than model size alone.
Interestingly, the Natural Language Probability remained relatively consistent across models (0.00013-0.0017), indicating that the models maintain a stable baseline for general language generation despite varying degrees of specialization for programming tasks. This observation provides a valuable reference point for understanding the models' relative biases toward programming-specific tokens.
When examining the Keyword Average Probability, we found that larger models did not necessarily exhibit higher probabilities for programming keywords. For instance, DeepSeek-R1-Distill-Qwen-1.5B showed a higher average probability (1.53E-05) compared to its larger counterpart DeepSeek-R1-Distill-Qwen-7B (2.78E-07), suggesting that model size may not be the primary factor in determining programming token representation.

\subsubsection{Top Token Analysis}
Further investigation of the most probable tokens in each category revealed distinct patterns across models. In programming keywords, commonly used terms like 'import', 'def', and 'void' consistently appeared among the top-3 keywords, though their relative rankings varied between models. This suggests a common foundation in basic programming constructs while maintaining model-specific specializations.
The analysis of top special tokens revealed particularly interesting patterns across the model families. The DeepSeek-R1-Distill-Qwen-1.5B showed a strong preference for basic syntax elements, with the newline character being its most probable special token. This suggests a fundamental understanding of code formatting structure. In contrast, the larger DeepSeek-R1-Distill-Qwen-7B favored XML-style tokens, potentially indicating a broader exposure to markup languages during training.
Interestingly, model size appeared to influence the complexity of preferred special tokens. The DeepSeek-R1-Distill-Qwen-14B favored the C-style format specifier, while the 32B variant showed a preference for basic parentheses. The DeepSeek-R1-Distill-Llama-8B exhibited a unique pattern with the hash symbol as its top special token, suggesting possible optimization for Python-style comments or preprocessing directives.
This variation in top special tokens across models provides insight into their architectural biases and potential specializations:

Syntax-focused models (e.g., DeepSeek-R1-Distill-Qwen-1.5B) with high probabilities for basic code formatting tokens
Markup-oriented models (e.g., DeepSeek-R1-Distill-Qwen-7B) showing preference for XML/HTML-style tokens
Language-specific models (e.g., DeepSeek-R1-Distill-Llama-8B) with special tokens commonly used in specific programming languages

The correlation between top programming keywords and special tokens further reinforces these specialization patterns. For instance, models with 'def' and 'import' among their top keywords often showed higher probabilities for Python-related special tokens, suggesting consistent language-specific training effects.

\subsection{The Ultimate Pythonist's Question: Tabs vs. Spaces}
Beyond the traditional evaluation metrics, we explored a unique aspect of code generation: the models' inherent preferences for different indentation styles. This analysis offers an empirical perspective on the famous "tabs versus spaces" debate in Python programming.
Our cold start probability analysis reveals several intriguing patterns (Appendix Table~\ref{format-token-probabilities}). First, the smallest model (DeepSeek-R1-Distill-Qwen-1.5B) shows an overwhelming preference for newlines (6.70E-02) over any indentation method, suggesting a focus on basic code structure rather than specific formatting. In contrast, larger models demonstrate more nuanced preferences, with the 32B variant showing relatively balanced probabilities between different indentation styles.
Remarkably, across all models, spaces consistently outperform tabs in probability, with four-space indentation being particularly favored in larger models. The DeepSeek-R1-Distill-Llama-8B shows the most dramatic space-over-tab preference, with four-space indentation (8.71E-05) being approximately 60 times more probable than tabs (1.43E-06). This aligns with PEP 8 Python style guidelines, suggesting that these models have naturally absorbed common coding conventions during training.
This finding adds an interesting dimension to our understanding of how models learn and represent programming conventions, showing that even without explicit programming, they develop clear preferences that mirror real-world coding practices.

\subsection{Compression Impact}

Next, we decided to focus on three key approaches commonly observed in the production deployment of models, beyond differences between architectures. Instead of comparing different models, we examined how a single family of models changes under three distinct processes:

\begin{enumerate}
    \item Quantization: This process reduces the precision and dimensionality of numerical representations in model weights, effectively compressing the model while maintaining performance.
    \item Model size: We examined models of different sizes (1.5B, 7B, 14B, and 32B) to assess how scale impacts performance and efficiency.
    \item Model type: We considered three key variants—base models, instruct-tuned models, and code models - to evaluate their respective optimizations and intended use cases.
\end{enumerate}

For our investigation, we focused exclusively on  Qwen2.5 family of models, as they are open-source and provide well-structured series of models that vary systematically both in quantization levels and distillation stages. This allowed us to systematically analyze the effects of these transformations on programming language token representations (Appendix Table~\ref{tokenizer-comparison}).

\subsubsection{Quantization Effects}
For a deeper understanding of how model compression affects coding abilities, we analyzed various quantization levels of the Qwen2.5-coder-7B model. It is interesting to note that the quantized Q8\_0 model is very close to the original model. Our analysis revealed several critical patterns that challenge initial assumptions about model performance.
A key finding emerged in the distribution between programming keywords and special tokens. While Q2\_K showed a concerning ratio with high special token probability (0.4248) compared to keyword probability (0.0902), less aggressive quantization actually improved this balance. The Q4\_K\_S variant, for instance, demonstrated a more favorable distribution with reduced special token probability (0.2804) and increased keyword probability (0.1231). This observation aligns with recent findings on quantization robustness \citep{fang2025smallerweakerbenchmarking}, which demonstrate that quantized models can exhibit superior resilience to noise perturbations compared to their full-precision counterparts.
This shift is particularly significant because high special token probabilities indicate a model's tendency to generate noise – repeated spaces, brackets, hashes, and other non-semantic elements that can lead to degenerate output patterns. For example, we observed tokens representing multiple repeated spaces, which could cause the model to get stuck in loops of meaningless character generation.
The consistency in top programming keywords ('import', 'package', 'def', 'from') across quantization levels suggests that essential coding semantics are preserved. However, the reduced probability of special tokens in more aggressive quantization schemes might actually represent an improvement in generation quality, as it indicates a lower likelihood of producing syntactic noise.

\subsection{Impact of Model Size and Instruction Tuning}
Our analysis of the Qwen2.5 model family across different sizes (1.5B to 32B) and variants (base, instruct, and coder) revealed several critical patterns in code generation capabilities. The data presents a complex interplay between model size, specialization, and instruction tuning.
\subsubsection{Base Model Scaling}
In the base Qwen2.5 models, we observed a relatively stable pattern of keyword probabilities across different sizes (0.143 for 1.5B, 0.113 for 7B, 0.123 for 14B, and 0.142 for 32B). Notably, these models maintained a healthier ratio between programming keywords and special tokens, with special token probabilities remaining consistently lower (around 0.12-0.15). This suggests that base models develop a more balanced token distribution regardless of size.
\subsubsection{Instruction Tuning Effects}
A particularly interesting finding emerged in the impact of instruction tuning. Across all sizes, instruction-tuned variants showed a concerning trend toward increased special token probabilities. This effect was most visible in the 7B-Coder-Instruct model, where special token probability reached 0.449 – potentially indicating an increased risk of generating syntactic noise.
The most striking degradation appeared in smaller instruction-tuned models. For instance, the Qwen2.5-Coder-1.5B-Instruct showed a drop in keyword probability (0.0003) compared to its base version (0.121), while maintaining a relatively high special token probability. This suggests that instruction tuning might be particularly challenging for smaller models to maintain code generation quality. The most significant difference we observe, aside from the fact that the Coder 1.5-Instruct model differs the most, is that it has completely different keywords. It’s unclear what happened to it, but it stands out considerably. In other words, there is one model where entirely different keywords suddenly appear at the top of the output. Additionally, its Keywords Probability is significantly lower.

\subsubsection{Size-Dependent Patterns}

The base 32B model maintained strong keyword probabilities (0.142) with balanced special token distribution
The Coder-32B version showed the highest keyword probability (0.173) among all variants
However, its instruction-tuned version (Coder-32B-Instruct) exhibited a significant degradation (0.027), suggesting that even large models aren't immune to instruction tuning artifacts. These findings suggest that while larger models generally handle code generation better, instruction tuning can introduce unexpected biases toward special token generation, potentially compromising code generation quality regardless of model size.

\subsection{Difference with distilled models}

The comparison between original Qwen models and their distilled DeepSeek-R1 counterparts reveals striking differences in token probability distributions, suggesting significant transformations in model behavior through the distillation process.
Most notably, the distilled models show  lower programming keyword probabilities compared to their original counterparts. For instance, while Qwen2.5-1.5B maintains a healthy keyword probability of 0.143, its distilled version (DeepSeek-R1-Distill-Qwen-1.5B) drops to merely 0.0042 – a reduction of nearly 97\%. This pattern persists across all model sizes, with the 32B variant showing a similar decrease from 0.142 to 0.0019.
Even more concerning is the shift in special token probabilities. The distilled 1.5B model shows an increase in special token probability (0.7246) compared to its original version (0.125), suggesting a strong bias toward generating syntactic noise. This indicates that the distillation process might be inadvertently amplifying the model's tendency to produce non-semantic tokens while diminishing its ability to generate meaningful programming constructs.
The transformation is also evident in the models' preferred tokens. While original Qwen-family models consistently favor semantic programming keywords ('import', 'package', 'from'), the distilled versions show a shift toward more basic constructs ('void', 'int', 'is') and exhibit higher probabilities for formatting-related special tokens. This suggests that the distillation process, while potentially optimizing for model size and speed, might be compromising the models' ability to generate coherent code. This degradation mirrors the theoretical findings \citep{inproceedings} regarding 'model collapse'. Their analysis demonstrates that models trained on synthetic outputs (as in the distillation process) tend to lose variance in their probability distributions, effectively pruning the vocabulary tails and converging onto a narrow set of high-frequency tokens.
It raises important considerations about the trade-offs involved in model distillation, particularly for code generation tasks where maintaining a balance between semantic and syntactic elements is crucial.

\subsection{Temperature effect}

An interesting aspect of text generation is that the raw logits produced by a model can differ significantly from what the end user actually sees. This discrepancy arises due to post-processing techniques and sampling parameters, particularly temperature scaling, which can be implemented quite differently across models.

Temperature controls the level of randomness in token selection—lower values make the model more deterministic, while higher values increase diversity in output. However, different models may implement temperature scaling in non-uniform ways, leading to variations in how tokens are sampled even when given the same logits.

\section{Conclusions}

In this work, we present the first systematic token-level analysis of Large Language Models' behavior in code generation tasks, with particular emphasis on the effects of compression techniques such as quantization and distillation on programming construct representations. Our results make several significant contributions to the field of code-generating language models.
We introduce a novel analytical framework that includes a "cold start" probability metric and a comprehensive set of quantitative measures for evaluating model behavior. These metrics - Programming Keywords Probability, Special Tokens Probability, Keyword Average Probability, and Natural Language Probability - provide new insights into models' code generation capabilities without requiring extensive testing. 

Through systematic application of these metrics, we demonstrate that token probability distributions at initialization can serve as reliable predictors of a model's coding abilities.
Our detailed characterization of tokenizer behavior across popular programming languages reveals several unexpected findings. Most notably, we discover that code-specialized models do not develop distinct token distributions compared to general language models, challenging common assumptions about model specialization. Furthermore, we identify significant differences in token-level behavior between models even when they share common vocabularies, suggesting that architectural choices and training objectives may have a stronger influence on coding capabilities than previously understood.

Through extensive experimentation with compression techniques, we uncover non-linear effects in model behavior. Particularly noteworthy is our finding that moderate quantization can actually improve the balance between programming keywords and special tokens, while more aggressive compression through distillation significantly alters token probability distributions, reducing programming keyword probabilities by up to 97\% while increasing special token probabilities. We also identify concerning patterns in instruction-tuned models, particularly in smaller variants, where tuning can lead to decreased programming keyword probabilities.
These findings not only advance our theoretical understanding of token-level mechanisms in code-generating language models but also provide practical guidelines for deploying compressed models in production environments. Our work establishes clear metrics and methodologies for evaluating the trade-offs between model compression and code generation capabilities, enabling more informed decisions in resource-constrained environments while maintaining reliable code generation quality.

\bibliography{iclr2025_conference}
\bibliographystyle{iclr2025_conference}

\appendix
\section{Appendix}

\begin{longtable}{lrrrrlll}
\caption{Keyword Token Rank Analysis. Columns include number of languages using each keyword, minimum rank, median rank, and ranks across tokenizers (Qwen2.5, DeepSeek-V3, Llama 3.1).}
\label{keyword-token-rank-analysis}\\
\toprule
   \bf Keyword &  \bf \# Langs &  \bf Min Rank &  \bf Median Rank &  \bf \# Tokenizers & \bf Qwen2.5 & \bf DeepSeek-V3 & \bf Llama \\
\midrule
\endfirsthead
\caption[]{Keyword Token Rank Analysis. Columns include number of languages using each keyword, minimum rank, median rank, and ranks across tokenizers (Qwen2.5, DeepSeek-V3, Llama).} \\
\toprule
   \bf Keyword &  \bf \# Langs &  \bf Min Rank &  \bf Median Rank &  \bf \# Tokenizers & \bf Qwen2.5 & \bf DeepSeek-V3 & \bf Llama \\
\midrule
\endhead
\midrule
\multicolumn{8}{r}{{Continued on next page}} \\
\midrule
\endfoot

\bottomrule
\endlastfoot
        in &              6 &     258.0 &        258.0 &               3 &      258 &         261 &   258 \\
        or &              4 &     269.0 &        269.0 &               3 &      269 &         272 &   269 \\
        is &              2 &     278.0 &        285.0 &               3 &      285 &         278 &   285 \\
        as &              5 &     300.0 &        300.0 &               3 &      300 &         306 &   300 \\
        if &             11 &     333.0 &        333.0 &               3 &      333 &         394 &   333 \\
       int &              4 &     396.0 &        396.0 &               3 &      396 &         650 &   396 \\
       end &              1 &     408.0 &        408.0 &               3 &      408 &         523 &   408 \\
       out &              1 &     411.0 &        412.0 &               3 &      411 &         606 &   412 \\
       and &              4 &     437.0 &        438.0 &               3 &      437 &         458 &   438 \\
    import &              5 &     474.0 &        475.0 &               3 &      474 &        1897 &   475 \\
      this &              5 &     574.0 &        576.0 &               3 &      574 &        3779 &   576 \\
    return &             11 &     689.0 &        693.0 &               3 &      689 &        3916 &   693 \\
      self &              2 &     721.0 &        726.0 &               3 &      721 &        2161 &   726 \\
       def &              2 &     750.0 &        755.0 &               3 &      750 &        3465 &   755 \\
       key &              1 &     792.0 &        798.0 &               3 &      792 &        4989 &   798 \\
       use &              2 &     810.0 &        817.0 &               3 &      810 &        3103 &   817 \\
    public &              6 &     888.0 &        898.0 &               3 &      888 &        3978 &   898 \\
    string &              1 &     917.0 &        928.0 &               3 &      917 &        4463 &   928 \\
       new &              6 &     931.0 &        943.0 &               3 &      931 &        2839 &   943 \\
       var &              4 &     947.0 &        959.0 &               3 &      947 &        5241 &   959 \\
     using &              2 &     970.0 &        985.0 &               3 &      970 &        4079 &   985 \\
   include &              1 &     997.0 &       1012.0 &               3 &      997 &        5211 &  1012 \\
      void &              6 &    1004.0 &       1019.0 &               3 &     1004 &        6483 &  1019 \\
      lock &              1 &    1023.0 &       1039.0 &               3 &     1023 &       10566 &  1039 \\
     const &              8 &    1024.0 &       1040.0 &               3 &     1024 &        3949 &  1040 \\
     class &              8 &    1040.0 &       1058.0 &               3 &     1040 &        3767 &  1058 \\
       ref &              3 &    1097.0 &       1116.0 &               3 &     1097 &        5044 &  1116 \\
       let &              3 &    1149.0 &       1169.0 &               3 &     1149 &        1775 &  1169 \\
    struct &              5 &    1235.0 &       1257.0 &               3 &     1235 &        8325 &  1257 \\
      type &              2 &    1313.0 &       1337.0 &               3 &     1313 &        4611 &  1337 \\
     print &              1 &    1350.0 &       1374.0 &               3 &     1350 &        3098 &  1374 \\
      from &              1 &    1499.0 &       1527.0 &               3 &     1499 &        5356 &  1527 \\
      else &             11 &    1503.0 &       1531.0 &               3 &     1503 &        9267 &  1531 \\
    export &              3 &    1533.0 &       1562.0 &               3 &     1533 &        7449 &  1562 \\
       try &              8 &    1539.0 &       1568.0 &               3 &     1539 &       26976 &  1568 \\
  function &              3 &    1688.0 &       1723.0 &               3 &     1688 &        8701 &  1723 \\
    object &              1 &    1700.0 &       1735.0 &               3 &     1700 &       10325 &  1735 \\
   package &              4 &    1722.0 &       1757.0 &               3 &     1722 &        7249 &  1757 \\
    select &              1 &    1742.0 &       1779.0 &               3 &     1742 &       13933 &  1779 \\
      char &              4 &    1762.0 &       1799.0 &               3 &     1762 &        7526 &  1799 \\
      true &              7 &    1866.0 &       1904.0 &               3 &     1866 &       11476 &  1904 \\
       not &              3 &    1921.0 &       1962.0 &               3 &     1921 &        2869 &  1962 \\
       for &             11 &    1958.0 &       2000.0 &               3 &     1958 &        2251 &  2000 \\
    static &              8 &    1978.0 &       2020.0 &               3 &     1978 &       15514 &  2020 \\
   private &              6 &    1996.0 &       2039.0 &               3 &     1996 &       12339 &  2039 \\
       box &              1 &    2011.0 &       2054.0 &               3 &     2011 &        7353 &  2054 \\
   context &              1 &    2147.0 &       2196.0 &               3 &     2147 &       15689 &  2196 \\
       map &              1 &    2186.0 &       2235.0 &               3 &     2186 &       10865 &  2235 \\
 Component &              1 &    2189.0 &       2238.0 &               3 &     2189 &       10220 &  2238 \\
    assert &              2 &    2207.0 &       2256.0 &               3 &     2207 &       12609 &  2256 \\
    signed &              2 &    2215.0 &       2239.5 &               2 &     2215 &           - &  2264 \\
 namespace &              3 &    2231.0 &       2280.0 &               3 &     2231 &       10972 &  2280 \\
   default &              8 &    2258.0 &       2309.0 &               3 &     2258 &       14979 &  2309 \\
     endif &              1 &    2330.0 &       2384.0 &               3 &     2330 &       26297 &  2384 \\
     state &              1 &    2454.0 &       2513.0 &               3 &     2454 &        9395 &  2513 \\
      uint &              1 &    2496.0 &       2557.0 &               3 &     2496 &       34671 &  2557 \\
      True &              1 &    2514.0 &       2575.0 &               3 &     2514 &       10634 &  2575 \\
       mod &              1 &    2593.0 &       2658.0 &               3 &     2593 &        5158 &  2658 \\
      bool &              3 &    2641.0 &       2707.0 &               3 &     2641 &       20600 &  2707 \\
        go &              1 &    2710.0 &       3346.0 &               3 &     3346 &        2710 &  3427 \\
      func &              1 &    2830.0 &       2900.0 &               3 &     2830 &       12642 &  2900 \\
     where &              1 &    2870.0 &       2940.0 &               3 &     2870 &        4779 &  2940 \\
       NaN &              4 &    2921.0 &       2994.0 &               3 &     2921 &       16042 &  2994 \\
        do &              9 &    2982.0 &       3055.0 &               3 &     2982 &        4016 &  3055 \\
      echo &              1 &    3047.0 &       3123.0 &               3 &     3047 &       18843 &  3123 \\
     event &              1 &    3087.0 &       3163.0 &               3 &     3087 &       24875 &  3163 \\
      base &              1 &    3152.0 &       3231.0 &               3 &     3152 &       14707 &  3231 \\
      then &              1 &    3391.0 &       3473.0 &               3 &     3391 &        9594 &  3473 \\
      move &              1 &    3397.0 &       3479.0 &               3 &     3397 &       47632 &  3479 \\
     while &             10 &    3472.0 &       3556.0 &               3 &     3472 &        8848 &  3556 \\
    params &              1 &    3519.0 &       3603.0 &               3 &     3519 &       20579 &  3603 \\
      next &              1 &    3600.0 &       3684.0 &               3 &     3600 &        6695 &  3684 \\
  unsigned &              2 &    3626.0 &       3710.0 &               3 &     3626 &       35536 &  3710 \\
     float &              4 &    3649.0 &       3733.0 &               3 &     3649 &       15891 &  3733 \\
      byte &              2 &    3782.0 &       3867.0 &               3 &     3782 &       23535 &  3867 \\
     false &              7 &    3849.0 &       3934.0 &               3 &     3849 &       19836 &  3934 \\
      auto &              2 &    3902.0 &       3989.0 &               3 &     3902 &       18414 &  3989 \\
      chan &              1 &    3991.0 &       5658.0 &               3 &     5658 &        3991 &  5776 \\
     False &              1 &    4049.0 &       4139.0 &               3 &     4049 &       15444 &  4139 \\
      None &              1 &    4064.0 &       4155.0 &               3 &     4064 &       19906 &  4155 \\
      with &              3 &    4197.0 &       4291.0 &               3 &     4197 &        6135 &  4291 \\
  template &              1 &    4214.0 &       4308.0 &               3 &     4214 &       27793 &  4308 \\
    extern &              4 &    4301.0 &       4399.0 &               3 &     4301 &       76215 &  4399 \\
   require &              1 &    4310.0 &       4408.0 &               3 &     4310 &       22716 &  4408 \\
    double &              4 &    4331.0 &       4429.0 &               3 &     4331 &       20563 &  4429 \\
    module &              1 &    4352.0 &       4450.0 &               3 &     4352 &       20038 &  4450 \\
    delete &              3 &    4542.0 &       4644.0 &               3 &     4542 &       23565 &  4644 \\
       END &              1 &    4689.0 &       4794.0 &               3 &     4689 &       16860 &  4794 \\
     props &              1 &    4761.0 &       4866.0 &               3 &     4761 &       27032 &  4866 \\
      long &              4 &    4825.0 &       4930.0 &               3 &     4825 &        9938 &  4930 \\
 interface &              6 &    4970.0 &       5077.0 &               3 &     4970 &       23964 &  5077 \\
    inline &              2 &    5057.0 &       5167.0 &               3 &     5057 &       46900 &  5167 \\
   typedef &              2 &    5286.0 &       5399.0 &               3 &     5286 &       29218 &  5399 \\
      case &              9 &    5638.0 &       5756.0 &               3 &     5638 &       11675 &  5756 \\
 protected &              6 &    5764.0 &       5883.0 &               3 &     5764 &       40252 &  5883 \\
  children &              1 &    5864.0 &       5988.0 &               3 &     5864 &       20434 &  5988 \\
   boolean &              1 &    6117.0 &       6245.0 &               3 &     6117 &       36541 &  6245 \\
  register &              2 &    6343.0 &       6477.0 &               3 &     6343 &       28766 &  6477 \\
     match &              2 &    6347.0 &       6481.0 &               3 &     6347 &       50374 &  6481 \\
      impl &              1 &    6383.0 &       6517.0 &               3 &     6383 &       31281 &  6517 \\
      pass &              1 &    6385.0 &       6519.0 &               3 &     6385 &        9762 &  6519 \\
       mut &              1 &    6984.0 &       7129.0 &               3 &     6984 &       30293 &  7129 \\
     throw &              6 &    7119.0 &       7265.0 &               3 &     7119 &       42117 &  7265 \\
     catch &              6 &    7173.0 &       7320.0 &               3 &     7173 &       41163 &  7320 \\
     begin &              1 &    7265.0 &       7413.0 &               3 &     7265 &        8277 &  7413 \\
    render &              1 &    7322.0 &       7472.0 &               3 &     7322 &       28324 &  7472 \\
   checked &              1 &    7549.0 &       7702.0 &               3 &     7549 &       47569 &  7702 \\
     async &              2 &    7692.0 &       7847.0 &               3 &     7692 &       41072 &  7847 \\
  operator &              2 &    7884.0 &       8043.0 &               3 &     7884 &       20178 &  8043 \\
    throws &              1 &    8100.0 &       8262.0 &               3 &     8100 &       61101 &  8262 \\
       nil &              1 &    8385.0 &       8551.0 &               3 &     8385 &       68782 &  8551 \\
     short &              4 &    8676.0 &       8846.0 &               3 &     8676 &       31150 &  8846 \\
   foreach &              2 &    8808.0 &       8984.0 &               3 &     8808 &       41810 &  8984 \\
        fn &              2 &    8822.0 &       8998.0 &               3 &     8822 &       28952 &  8998 \\
     break &             11 &    8960.0 &       9137.0 &               3 &     8960 &       15485 &  9137 \\
      enum &              7 &    9018.0 &       9195.0 &               3 &     9018 &       47002 &  9195 \\
  typename &              1 &    9031.0 &       9208.0 &               3 &     9031 &       50578 &  9208 \\
  override &              2 &    9199.0 &       9380.0 &               3 &     9199 &       44462 &  9380 \\
      when &              1 &    9309.0 &       9493.0 &               3 &     9309 &       21123 &  9493 \\
     super &              5 &    9522.0 &       9712.0 &               3 &     9522 &       27196 &  9712 \\
  continue &             10 &    9534.0 &       9726.0 &               3 &     9534 &       45911 &  9726 \\
       pub &              1 &    9585.0 &       9780.0 &               3 &     9585 &       38744 &  9780 \\
       del &              1 &    9588.0 &       9783.0 &               3 &     9588 &       18421 &  9783 \\
     range &              1 &    9669.0 &       9866.0 &               3 &     9669 &       16801 &  9866 \\
    global &              2 &    9752.0 &       9951.0 &               3 &     9752 &       31754 &  9951 \\
    typeof &              5 &   10222.0 &      10433.0 &               3 &    10222 &       73890 & 10433 \\
    sizeof &              3 &   10318.0 &      10531.0 &               3 &    10318 &       58850 & 10531 \\
  internal &              1 &   10481.0 &      10701.0 &               3 &    10481 &       52430 & 10701 \\
      loop &              1 &   10498.0 &      10719.0 &               3 &    10498 &       42938 & 10719 \\
       asm &              1 &   10530.0 &      10753.0 &               3 &    10530 &       18969 & 10753 \\
    friend &              1 &   10701.0 &      10931.0 &               3 &    10701 &       26736 & 10931 \\
     await &              3 &   11421.0 &      11675.0 &               3 &    11421 &       62743 & 11675 \\
    except &              1 &   11683.0 &      11945.0 &               3 &    11683 &       32812 & 11945 \\
     final &              4 &   11822.0 &      12085.0 &               3 &    11822 &       41551 & 12085 \\
      priv &              1 &   11887.0 &      12151.0 &               3 &    11887 &       78737 & 12151 \\
      Self &              1 &   12092.0 &      12363.0 &               3 &    12092 &       37400 & 12363 \\
      elif &              1 &   12458.0 &      12740.0 &               3 &    12458 &       53358 & 12740 \\
    lambda &              1 &   12935.0 &      13077.0 &               3 &    12935 &       13077 & 13231 \\
     React &              1 &   14799.0 &      15143.0 &               3 &    14799 &       54635 & 15143 \\
     alias &              1 &   14956.0 &      15305.0 &               3 &    14956 &       88739 & 15305 \\
   extends &              4 &   15231.0 &      15588.0 &               3 &    15231 &       83878 & 15588 \\
     undef &              1 &   16135.0 &      16331.0 &               2 &    16135 &           - & 16527 \\
     union &              2 &   16192.0 &      16588.0 &               3 &    16192 &       32099 & 16588 \\
  abstract &              4 &   16249.0 &      16647.0 &               3 &    16249 &       42149 & 16647 \\
    switch &              8 &   17338.0 &      17790.0 &               3 &    17338 &       46941 & 17790 \\
   declare &              1 &   18471.0 &      18978.0 &               3 &    18471 &       73780 & 18978 \\
     raise &              1 &   18704.0 &      19223.0 &               3 &    18704 &       55003 & 19223 \\
  provider &              1 &   19979.0 &      20576.0 &               3 &    19979 &       96058 & 20576 \\
     clone &              1 &   19982.0 &      20579.0 &               3 &    19982 &      102131 & 20579 \\
      hook &              1 &   20873.0 &      21543.0 &               3 &    20873 &       80513 & 21543 \\
     fixed &              1 &   22021.0 &      22795.0 &               3 &    22021 &       77769 & 22795 \\
endforeach &              1 &   22095.0 &      22871.0 &               3 &    22095 &      122655 & 22871 \\
  readonly &              2 &   22569.0 &      22977.0 &               2 &    22569 &           - & 23385 \\
   decimal &              1 &   23289.0 &      24170.0 &               3 &    23289 &      103814 & 24170 \\
   virtual &              3 &   25668.0 &      26752.0 &               3 &    25668 &       76356 & 26752 \\
    elseif &              1 &   26016.0 &      26561.5 &               2 &    26016 &           - & 27107 \\
    ensure &              1 &   27289.0 &      28389.0 &               3 &    27289 &       94747 & 28389 \\
  volatile &              4 &   27307.0 &      27857.0 &               2 &    27307 &           - & 28407 \\
  restrict &              1 &   27898.0 &      28448.0 &               2 &    27898 &           - & 28998 \\
  delegate &              1 &   28227.0 &      28777.0 &               2 &    28227 &           - & 29327 \\
      goto &              5 &   28535.0 &      29635.0 &               3 &    28535 &      105723 & 29635 \\
  \_Generic &              1 &   29085.0 &      29635.0 &               2 &    29085 &           - & 30185 \\
     trait &              2 &   29432.0 &      29982.0 &               2 &    29432 &           - & 30532 \\
     yield &              6 &   29696.0 &      30246.0 &               2 &    29696 &           - & 30796 \\
    native &              1 &   29738.0 &      30838.0 &               3 &    29738 &      109720 & 30838 \\
   mutable &              1 &   30473.0 &      31023.0 &               2 &    30473 &           - & 31573 \\
  implicit &              1 &   30940.0 &      31490.0 &               2 &    30940 &           - & 32040 \\
 unchecked &              1 &   31684.0 &      32234.0 &               2 &    31684 &           - & 32784 \\
     macro &              1 &   32606.0 &      33706.0 &               3 &    32606 &      116634 & 33706 \\
     ulong &              1 &   32832.0 &      33382.0 &               2 &    32832 &           - & 33932 \\
     BEGIN &              1 &   37588.0 &      38688.0 &               3 &    37588 &      104472 & 38688 \\
    portal &              1 &   37953.0 &      39053.0 &               3 &    37953 &      104008 & 39053 \\
    unsafe &              2 &   38157.0 &      38707.0 &               2 &    38157 &           - & 39257 \\
     until &              1 &   38730.0 &      39830.0 &               3 &    38730 &       56657 & 39830 \\
    unless &              1 &   38770.0 &      39870.0 &               3 &    38770 &       69190 & 39870 \\
   finally &              6 &   39176.0 &      40276.0 &               3 &    39176 &      121229 & 40276 \\
    ushort &              1 &   40375.0 &      40925.0 &               2 &    40375 &           - & 41475 \\
   nullptr &              2 &   41132.0 &      41682.0 &               2 &    41132 &           - & 42232 \\
  requires &              1 &   41375.0 &      42475.0 &               3 &    41375 &      101936 & 42475 \\
  fragment &              1 &   42202.0 &      42752.0 &               2 &    42202 &           - & 43302 \\
 constexpr &              2 &   42281.0 &      42831.0 &               2 &    42281 &           - & 43381 \\
       dyn &              1 &   43085.0 &      43635.0 &               2 &    43085 &           - & 44185 \\
     retry &              1 &   44848.0 &      45398.0 &               2 &    44848 &           - & 45948 \\
       jsx &              1 &   45290.0 &      45840.0 &               2 &    45290 &           - & 46390 \\
  consumer &              1 &   46764.0 &      47864.0 &               3 &    46764 &      114495 & 47864 \\
      lazy &              1 &   49013.0 &      49563.0 &               2 &    49013 &           - & 50113 \\
  useState &              1 &   55670.0 &      56220.0 &               2 &    55670 &           - & 56770 \\
     crate &              1 &   61711.0 &      62261.0 &               2 &    61711 &           - & 62811 \\
     defer &              1 &   62095.0 &      62645.0 &               2 &    62095 &           - & 63195 \\
      redo &              1 &   63561.0 &      64661.0 &               3 &    63561 &      127260 & 64661 \\
     elsif &              1 &   65967.0 &      66517.0 &               2 &    65967 &           - & 67067 \\
   concept &              1 &   68487.0 &      69037.0 &               2 &    68487 &           - & 69587 \\
       xor &              2 &   71311.0 &      72411.0 &               3 &    71311 &      114718 & 72411 \\
  decltype &              1 &   74364.0 &      74914.0 &               2 &    74364 &           - & 75464 \\
    sealed &              2 &   75940.0 &      76490.0 &               2 &    75940 &           - & 77040 \\
     \_Bool &              1 &   79948.0 &      80498.0 &               2 &    79948 &           - & 81048 \\
    typeid &              1 &   88342.0 &      88892.0 &               2 &    88342 &           - & 89442 \\
     compl &              1 &   92843.0 &      92843.0 &               1 &        - &       92843 &     - \\
  explicit &              2 &   93632.0 &      94182.0 &               2 &    93632 &           - & 94732 \\
implements &              4 &   93958.0 &      94508.0 &               2 &    93958 &           - & 95058 \\
  callable &              1 &   95192.0 &      95742.0 &               2 &    95192 &           - & 96292 \\
\end{longtable}

\begin{table}[ht]
\caption{Coverage of Programming Language Keywords in Tokenizer Vocabularies. A comparison of how well different tokenizers represent reserved keywords from various programming languages. Percentages indicate the proportion of keywords found in each tokenizer’s vocabulary.}
\label{keyword-coverage-comparison}
\begin{center}
\begin{tabular}{lrrrr}
\textbf{Language} & \textbf{Keywords} & \textbf{Qwen2.5-Coder (\%)} & \textbf{DeepSeek-R1 (\%)} & \textbf{Llama (\%)} \\
\hline
C             & 59  & 71.2 & 59.3 & 71.2 \\
C\#            & 77  & 97.4 & 84.4 & 97.4 \\
TypeScript    & 46  & 95.7 & 91.3 & 95.7 \\
Ruby          & 41  & 87.8 & 78.0 & 87.8 \\
PHP           & 62  & 85.5 & 75.8 & 85.5 \\
Rust          & 51  & 96.1 & 86.3 & 96.1 \\
JavaScript    & 46  & 95.7 & 91.3 & 95.7 \\
Java          & 51  & 90.2 & 84.3 & 90.2 \\
Python        & 35  & 97.1 & 94.3 & 97.1 \\
Go            & 25  & 96.0 & 92.0 & 96.0 \\
React         & 30  & 56.7 & 43.3 & 56.7 \\
C++           & 93  & 73.1 & 64.5 & 73.1 \\
\hline
\end{tabular}
\end{center}
A full list of absent keywords in tokens: \_Complex, static\_assert, thread\_local, \_Thread\_local, \_Imaginary, \_Alignas, \_Alignof, alignas, \_Decimal64, \_Decimal128, \_BitInt, \_Atomic, \_Static\_assert, typeof\_unqual, alignof, \_Noreturn, \_Decimal32, sbyte, stackalloc, debugger, instanceof, LINE, FILE, ENCODING, defined?, rescue, enddeclare, endfor, endswitch, endwhile, include\_once, insteadof, require\_once, yield from, become, unsized, permits, transient, strictfp, synchronized, nonlocal, fallthrough, reconciliation, useLayoutEffect, strictMode, useEffect, useImperativeHandle, errorBoundary, useCallback, suspense, useContext, useRef, useMemo, useReducer, virtualDOM, const\_cast, bitor, constinit, not\_eq, and\_eq, noexcept, char8\_t, xor\_eq, co\_yield, consteval, char16\_t, bitand, or\_eq, co\_await, reinterpret\_cast, wchar\_t, static\_cast, co\_return, dynamic\_cast, char32\_t
\end{table}

\begin{table}[ht]
\caption{Proportion of Tokens Containing Programming-Specific Special Characters in Tokenizer Vocabularies. The number and percentage of tokens that contain programming-related symbols across three major tokenizers.}
\label{special-chars-coverage}
\begin{center}
\begin{tabular}{lrrr}
\textbf{Tokenizer} & \textbf{Tokens with Special Chars} & \textbf{Total Tokens} & \textbf{Percentage (\%)} \\
\hline
Llama        & 18,719  & 128,000 & 14.6 \\
DeepSeek-V3 & 6,585   & 128,000 & 5.1  \\
Qwen2.5    & 18,454  & 151,643 & 12.1 \\
\hline
\end{tabular}
\end{center}
\end{table}

\begin{table*}[ht]
\caption{Comparison of Tokenizer Statistics Across Models. This table presents keyword probability, special token probability, and other relevant statistics for various tokenizers.}
\label{tokenizer-comparison-1}
\begin{center}
\rotatebox{90}{
\begin{tabular}{lrrrrrrll}
\textbf{Model} & \textbf{KeyW Prob} & \textbf{Spec tok Prob} & \textbf{KeyW Avg Prob} & \textbf{Spec tok Avg Prob} & \textbf{NL prob} & \textbf{Top-3 KeyW} & \textbf{Top-3 Spec} \\
\hline
R1-Qwen-1.5B & 0.0042 & 0.7246 & 1.53E-05 & 3.59E-05 & 0.00017 & map, void, int & )\textbackslash u010a\textbackslash u010a \\
R1-Qwen-7B & 0.0017 & 0.1046 & 2.78E-07 & 5.18E-06 & 0.00016 & void, public, private & \textbackslash u00ef\textbackslash u00bc\textbackslash u013c< \\
R1-Qwen-14B & 0.0011 & 0.1082 & 3.97E-06 & 5.36E-06 & 0.00013 & is, in, use & \%c \\
R1-Qwen-32B & 0.0019 & 0.2293 & 6.77E-06 & 1.14E-05 & 0.00038 & int, is, end & \textbackslash u0120( \\
R1-Llama-8B & 0.0062 & 0.1006 & 2.25E-05 & 4.98E-06 & 0.0004 & def, import, from & \# \\
\hline
\end{tabular}
}
\end{center}
\end{table*}

\begin{table}[ht]
\caption{Comparison of Formatting Token Probabilities Across Models. This table presents tabulation, new line, and space token probabilities for various tokenizers.}
\label{format-token-probabilities}
\begin{center}
\begin{tabular}{lrrrr}
\textbf{Model} & \textbf{Tab} & \textbf{New line} & \textbf{Two spaces} & \textbf{Four spaces} \\
\hline
R1-Qwen-1.5B & 3.58E-07 & 6.70E-02 & 4.47E-06 & 1.01E-06 \\
R1-Qwen-7B & 4.29E-06 & 5.62E-05 & 5.90E-06 & 4.53E-06 \\
R1-Qwen-14B & 2.09E-06 & 6.59E-05 & 9.18E-06 & 1.48E-05 \\
R1-Qwen-32B & 2.03E-06 & 0.04132080078 & 0.0002186 & 7.99E-05 \\
R1-Llama-8B & 1.43E-06 & 9.95E-05 & 1.75E-05 & 8.71E-05 \\
\hline
\end{tabular}
\end{center}
\end{table}

\begin{table*}[ht]
\caption{Comparison of Tokenizer Statistics Across Models. This table presents keyword probability, special token probability, and other relevant statistics for various tokenizers.}
\label{tokenizer-comparison}
\begin{center}
\rotatebox{90}{
\begin{tabular}{lrrrrrrll}
\textbf{Model} & \textbf{KeyW Prob} & \textbf{Spec tok Prob} & \textbf{KeyW Avg Prob} & \textbf{Spec tok Avg Prob} & \textbf{NL prob} & \textbf{Top-3 KeyW} & \textbf{Top-3 Spec} \\
\hline
Q2\_K & 0.0902 & 0.4248 & 0.0003 & 2.10E-05 & 0.0017 & import, package, def & **, \#, // \\
Q3\_K\_L & 0.1115 & 0.3289 & 0.0004 & 1.63E-05 & 0.0025 & import, package, from & **, \#, // \\
Q3\_K\_M & 0.1198 & 0.3320 & 0.0004 & 1.64E-05 & 0.0026 & import, package, def & **, \#, // \\
Q3\_K\_S & 0.0947 & 0.4147 & 0.0003 & 2.05E-05 & 0.0024 & import, package, def & **, \#, // \\
Q4\_0 & 0.0873 & 0.3353 & 0.0003 & 1.66E-05 & 0.0015 & import, package, const & **, \#, // \\
Q4\_1 & 0.0985 & 0.2821 & 0.0004 & 1.40E-05 & 0.0020 & import, package, def & **, \#, // \\
Q4\_K\_M & 0.1068 & 0.2900 & 0.0004 & 1.44E-05 & 0.0021 & import, package, from & **, \#, // \\
Q4\_K\_S & 0.1231 & 0.2804 & 0.0004 & 1.39E-05 & 0.0024 & import, package, def & **, \#, // \\
Q5\_0 & 0.1081 & 0.2830 & 0.0004 & 1.40E-05 & 0.0022 & import, package, from & **, \#, // \\
Q5\_1 & 0.1108 & 0.2867 & 0.0004 & 1.42E-05 & 0.0020 & import, package, from & **, \#, // \\
Q5\_K\_M & 0.1093 & 0.2929 & 0.0004 & 1.45E-05 & 0.0019 & import, package, from & **, \#, // \\
Q5\_K\_S & 0.1120 & 0.2928 & 0.0004 & 1.45E-05 & 0.0019 & import, package, from & **, \#, // \\
Q6\_K & 0.1083 & 0.3011 & 0.0004 & 1.49E-05 & 0.0020 & import, package, from & **, \#, // \\
Q8\_0 & 0.1070 & 0.3025 & 0.0004 & 1.50E-05 & 0.0019 & import, package, from & **, \#, // \\
Qwen2.5-1.5B & 0.1434 & 0.1254 & 0.0005 & 6.22E-06 & 0.0043 & import, package, from & \makecell[l]{\#, \#include, \\ $<$} \\
Qwen2.5-1.5B-Inst & 0.0389 & 0.4482 & 0.0001 & 2.22E-05 & 0.0040 & import, package, from & \#, \texttt{\textbackslash n}, $<$ \\
Qwen2.5-Coder-1.5B & 0.1215 & 0.3104 & 0.0004 & 1.54E-05 & 0.0005 & import, package, public & **, \#, /* \\
Qwen2.5-Coder-1.5B-Inst & 0.0003 & 0.0262 & 1.10E-06 & 1.30E-06 & 5.47E-05 & or, if, int & ., ,, / \\
Qwen2.5-7B & 0.1131 & 0.1470 & 0.0004 & 7.28E-06 & 0.0049 & import, public, from & \makecell[l]{\#, \#!/, \\ \#include} \\
Qwen2.5-7B-Inst & 0.1399 & 0.1644 & 0.0005 & 8.15E-06 & 0.0042 & import, public, def & \#, $<$, // \\
Qwen2.5-Coder-7B & 0.1080 & 0.2985 & 0.0004 & 1.48E-05 & 0.0019 & import, package, from & **, \#, // \\
Qwen2.5-Coder-7B-Inst & 0.2572 & 0.4495 & 0.0009 & 2.23E-05 & 0.0007 & import, package, const & **, \#, // \\
Qwen2.5-14B & 0.1232 & 0.1249 & 0.0004 & 6.19E-06 & 0.0068 & import, from, package & \#, \#!/, $<$ \\
Qwen2.5-14B-Inst & 0.0758 & 0.1062 & 0.0003 & 5.26E-06 & 0.0024 & import, package, public & \#, $<$, \#!/ \\
Qwen2.5-Coder-14B & 0.0857 & 0.2587 & 0.0003 & 1.28E-05 & 0.0019 & import, package, from & **, \#, // \\
Qwen2.5-Coder-14B-Inst & 0.0108 & 0.2361 & 3.91E-05 & 1.17E-05 & 0.0026 & package, import, public & /, -, : \\
\hline
\end{tabular}
}
\end{center}
\end{table*}


\end{document}

%% file: math_commands.tex
\usepackage{amsmath,amsfonts,bm}

\def\eqref#1{equation~\ref{#1}}

\def\1{\bm{1}}

\DeclareMathAlphabet{\mathsfit}{\encodingdefault}{\sfdefault}{m}{sl}
\SetMathAlphabet{\mathsfit}{bold}{\encodingdefault}{\sfdefault}{bx}{n}



%% file: iclr2025_conference.bib
@misc{chiang2024chatbot,
  title={Chatbot Arena: An Open Platform for Evaluating LLMs by Human Preference},
  author={Chiang, Wei-Lin and others},
  year={2024},
  eprint={2403.04132},
  archivePrefix={arXiv},
  primaryClass={cs.CL},
  url={https://doi.org/10.48550/arXiv.2403.04132},
  note={arXiv preprint}
}

@misc{deepseek2025deepseekr1,
  title={DeepSeek-R1: Incentivizing Reasoning Capability in LLMs via Reinforcement Learning},
  author={{DeepSeek-AI} and Guo, Daya and others},
  year={2025},
  eprint={2501.12948},
  archivePrefix={arXiv},
  primaryClass={cs.LG},
  url={https://doi.org/10.48550/arXiv.2501.12948},
  note={arXiv preprint}
}

@misc{deepseek2024deepseekv2,
  title={DeepSeek-V2: A Strong, Economical, and Efficient Mixture-of-Experts Language Model},
  author={{DeepSeek-AI} and Liu, Aixin and Feng, Bei and Wang, Bin and others},
  year={2024},
  eprint={2405.04434},
  archivePrefix={arXiv},
  primaryClass={cs.CL},
  url={https://doi.org/10.48550/arXiv.2405.04434},
  note={arXiv preprint}
}

@misc{deepseek2024deepseekv3,
  title={DeepSeek-V3 Technical Report},
  author={{DeepSeek-AI} and Liu, Aixin and Feng, Bei and Xue, Bing and others},
  year={2024},
  eprint={2412.19437},
  archivePrefix={arXiv},
  primaryClass={cs.CL},
  url={https://doi.org/10.48550/arXiv.2412.19437},
  note={arXiv preprint}
}

@misc{grattafiori2024llama3,
  title={The Llama 3 Herd of Models},
  author={Grattafiori, Aaron and others},
  year={2024},
  eprint={2407.21783},
  archivePrefix={arXiv},
  primaryClass={cs.CL},
  url={https://doi.org/10.48550/arXiv.2407.21783},
  note={arXiv preprint}
}

@misc{hui2024qwen25,
  title={Qwen2.5-Coder Technical Report},
  author={Hui, Binyuan and others},
  year={2024},
  eprint={2409.12186},
  archivePrefix={arXiv},
  primaryClass={cs.CL},
  url={https://doi.org/10.48550/arXiv.2409.12186},
  note={arXiv preprint}
}

@misc{nexusflow2024athenev2,
  title={Introducing Athene-V2: Advancing Beyond the Limits of Scaling with Targeted Post-Training},
  author={{Nexusflow}},
  year={2024},
  month={November},
  url={https://nexusflow.ai/blogs/athene-v2},
  note={Blog post}
}

@misc{githuttwo,
  author = {Fabian Beuke},
  title = {GitHut 2.0: GitHub Language Statistics},
  year = {2023},
  note = {GitHub repository},
  howpublished = {\url{https://madnight.github.io/githut/#/}}
}

@inproceedings{bostrom2020byte,
  title={Byte Pair Encoding Is Suboptimal for Language Model Pretraining},
  author={Bostrom, Kaj and Durrett, Greg},
  booktitle={Findings of the Association for Computational Linguistics: EMNLP 2020},
  publisher={Association for Computational Linguistics},
  year={2020},
  pages={4617--4624},
  doi={10.18653/v1/2020.findings-emnlp.414},
  url={https://doi.org/10.18653/v1/2020.findings-emnlp.414}
}

@misc{chen2021evaluating,
  title={Evaluating Large Language Models Trained on Code},
  author={Chen, Mark and others},
  year={2021},
  eprint={2107.03374},
  archivePrefix={arXiv},
  primaryClass={cs.LG},
  url={https://doi.org/10.48550/arXiv.2107.03374},
  note={arXiv preprint}
}

@misc{austin2021program,
  title={Program Synthesis with Large Language Models},
  author={Austin, Jacob and others},
  year={2021},
  eprint={2108.07732},
  archivePrefix={arXiv},
  primaryClass={cs.PL},
  url={https://doi.org/10.48550/arXiv.2108.07732},
  note={arXiv preprint}
}

@misc{ouyang2024lowbit,
  title={Low-Bit Quantization Favors Undertrained LLMs: Scaling Laws for Quantized LLMs with 100T Training Tokens},
  author={Ouyang, Xu and others},
  year={2024},
  eprint={2411.17691},
  archivePrefix={arXiv},
  primaryClass={cs.LG},
  url={https://doi.org/10.48550/arXiv.2411.17691},
  note={arXiv preprint}
}

@misc{dagan2024tokenizer,
  title={Getting the Most out of Your Tokenizer for Pre-Training and Domain Adaptation},
  author={Dagan, Gautier and others},
  year={2024},
  eprint={2402.01035},
  archivePrefix={arXiv},
  primaryClass={cs.CL},
  url={https://doi.org/10.48550/arXiv.2402.01035},
  note={arXiv preprint}
}

@misc{Chelombitko2024,
      title={Qtok: A Comprehensive Framework for Evaluating Multilingual Tokenizer Quality in Large Language Models}, 
      author={Iaroslav Chelombitko and Egor Safronov and Aleksey Komissarov},
      year={2024},
      eprint={2410.12989},
      archivePrefix={arXiv},
      primaryClass={cs.CL},
      url={https://doi.org/10.48550/arXiv.2410.12989}, 
      note={arXiv preprint}
}

@inproceedings{chelombitko-komissarov-2024-specialized,
    title = "Specialized Monolingual {BPE} Tokenizers for {U}ralic Languages Representation in Large Language Models",
    author = "Chelombitko, Iaroslav  and
      Komissarov, Aleksey",
    editor = {H{\"a}m{\"a}l{\"a}inen, Mika  and
      Pirinen, Flammie  and
      Macias, Melany  and
      Crespo Avila, Mario},
    booktitle = "Proceedings of the 9th International Workshop on Computational Linguistics for Uralic Languages",
    month = nov,
    year = "2024",
    address = "Helsinki, Finland",
    publisher = "Association for Computational Linguistics",
    url = "https://aclanthology.org/2024.iwclul-1.11/",
    pages = "89--95"
}

@misc{pagnoni2024bytelatenttransformerpatches,
      title={Byte Latent Transformer: Patches Scale Better Than Tokens}, 
      author={Artidoro Pagnoni and Ram Pasunuru and Pedro Rodriguez and John Nguyen and Benjamin Muller and Margaret Li and Chunting Zhou and Lili Yu and Jason Weston and Luke Zettlemoyer and Gargi Ghosh and Mike Lewis and Ari Holtzman and Srinivasan Iyer},
      year={2024},
      eprint={2412.09871},
      archivePrefix={arXiv},
      primaryClass={cs.CL},
      url={https://arxiv.org/abs/2412.09871}, 
}

@misc{zhuo2025bigcodebenchbenchmarkingcodegeneration,
      title={BigCodeBench: Benchmarking Code Generation with Diverse Function Calls and Complex Instructions}, 
      author={Terry Yue Zhuo and Minh Chien Vu and Jenny Chim and Han Hu and Wenhao Yu and Ratnadira Widyasari and Imam Nur Bani Yusuf and Haolan Zhan and Junda He and Indraneil Paul and Simon Brunner and Chen Gong and Thong Hoang and Armel Randy Zebaze and Xiaoheng Hong and Wen-Ding Li and Jean Kaddour and Ming Xu and Zhihan Zhang and Prateek Yadav and Naman Jain and Alex Gu and Zhoujun Cheng and Jiawei Liu and Qian Liu and Zijian Wang and Binyuan Hui and Niklas Muennighoff and David Lo and Daniel Fried and Xiaoning Du and Harm de Vries and Leandro Von Werra},
      year={2025},
      eprint={2406.15877},
      archivePrefix={arXiv},
      primaryClass={cs.SE},
      url={https://arxiv.org/abs/2406.15877}, 
}

@misc{jimenez2024swebenchlanguagemodelsresolve,
      title={SWE-bench: Can Language Models Resolve Real-World GitHub Issues?}, 
      author={Carlos E. Jimenez and John Yang and Alexander Wettig and Shunyu Yao and Kexin Pei and Ofir Press and Karthik Narasimhan},
      year={2024},
      eprint={2310.06770},
      archivePrefix={arXiv},
      primaryClass={cs.CL},
      url={https://arxiv.org/abs/2310.06770}, 
}

@misc{fang2025smallerweakerbenchmarking,
      title={Smaller = Weaker? Benchmarking Robustness of Quantized LLMs in Code Generation}, 
      author={Sen Fang and Weiyuan Ding and Antonio Mastropaolo and Bowen Xu},
      year={2025},
      eprint={2506.22776},
      archivePrefix={arXiv},
      primaryClass={cs.SE},
      url={https://arxiv.org/abs/2506.22776}, 
}

@inproceedings{inproceedings,
author = {Dohmatob, Elvis and Feng, Yunzhen and Kempe, Julia},
year = {2024},
month = {01},
pages = {46979-47013},
title = {Model Collapse Demystified: The Case of Regression},
doi = {10.52202/079017-1490}
}
